\documentclass[aps,prb,reprint,amsfonts,amsmath,showpacs]{revtex4-1}
\usepackage[english]{babel}
\usepackage[T1]{fontenc}
\usepackage[ansinew]{inputenc}
\usepackage{graphicx}
\usepackage{sistyle}
\usepackage{bm}
\usepackage{upgreek}

\begin{document}
\title{Josephson surface plasmons in spatially confined cuprate superconductors}
\author{Filippo Alpeggiani}
\email{filippo.alpeggiani01@ateneopv.it}
\author{Lucio Claudio Andreani}
\affiliation{Dipartimento di Fisica, Universit\`a di Pavia and CNISM, Via Bassi 6, 27100 Pavia, Italy}
\date{\today}

\begin{abstract}
In this work, we generalize the theory of localized surface plasmons to the case of high-$T_{\mathrm{c}}$ cuprate superconductors, spatially confined in the form of small spherical particles. At variance from ordinary metals, cuprate superconductors are characterized by a low-energy bulk excitation known as the Josephson plasma wave (JPW), arising from interlayer tunneling of the condensate along the $c$-axis. The effect of the JPW is revealed in a characteristic spectrum of surface excitations, which we call \emph{Josephson surface plasmons}. Our results, which apply to any material with a strongly anisotropic electromagnetic response, are worked out in detail for the case of multilayered superconductors supporting both low-frequency (acoustic) and transverse-optical JPW. Spatial confinement of the Josephson plasma waves may represent a new degree of freedom to engineer their frequencies and to explore the link between interlayer tunnelling and high-$T_{\mathrm{c}}$ superconductivity.
\end{abstract}

\maketitle

\section{Introduction}

The concept of spatial confinement in low-dimensional systems has led to a number of new ideas and achievements in several areas of nanophysics including semiconductors, photonics, and plasmonics. In particular, confined surface plasmons are widely exploited in the context of metallic nanoparticles \cite{maier,gaponenko,khurgin,review_plasmons,colas}. In this work, we generalize the same concept to the field of high-\Tc{} cuprate superconductors. These compounds are strongly anisotropic systems whose layered structure is composed of conductive CuO$_2$ planes stacked along the $c$-axis and separated by insulating block layers. As a consequence of their strong anisotropy, cuprate superconductors support two kinds of plasma excitations: the Josephson plasma wave (JPW), arising from interplane Josephson tunneling of the superfluid density \cite{bulaevskii1994time,helm2002opti,report}, and quasi-2D (acoustic) plasmons \cite{fetter, bill2003elec} that result from plasma oscillations along the planes, with noncondensed electrons being responsible for dissipation. The Josephson plasma wave propagates at the \emph{Josephson plasma frequency} $\wj$, which depends on the properties of the Josephson junction formed between two consecutive CuO$_2$ planes through the interplane distance and the critical current, thus providing a good estimate of the strength of interplane Josephson coupling.

In addition to bulk excitations, the study of \emph{surface} Josephson plasma waves propagating on planar dielectric--superconductor interfaces \cite{savelev2005surf,*yampolskii2008surf,*averkov2013obli} or in superconductor slabs \cite{doria1997plas, slipchenko2011surf} has been carried out. Surface modes in planar geometry cannot be directly excited by light, but can be probed with the attenuated-total-reflection method\cite{yampolskii2007exci,*yampolskii2009reso}. On the other hand, small particles made of cuprate superconductors present optical resonances whose frequencies can be related to the Josephson frequency. They are used to probe the Josephson coupling strength by an experimental technique called \emph{sphere resonance method}\cite{noh1990far,hirata2012corr}, which is commonly analyzed in the dipole approximation and in the framework of a simple Drude model. Here, a more complete theoretical framework for the interpretation of such experiments is presented.

The mechanisms determining the critical temperature \Tc{} of cuprate superconductors are not well established, yet \Tc{} is known to depend on the doping level, on apical-oxygen distance from the CuO$_2$ plane \cite{pavarini2001}, and in multilayered superconductors (MLSCs) it increases with the number $n$ of CuO$_2$ layers in a unit cell up to at least $n=3$ \cite{scott1994,plakida,hirata2012corr}. Studying the Josephson plasma wave may contribute to clarifying the mechanism of \highTc{} superconductivity, as it has been proposed that Josephson tunneling may be responsible for a lowering of the kinetic energy, i.e., an increase of the superconducting condensation energy\cite{anderson1995,leggett1999,munzar,molegraf2002supe}, which is proportional to $\wj^2$. This is especially likely to occur in multilayered superconductors, where inter- and intra-multilayer couplings give rise to both low-frequency (acoustic) and optical JPWs~\cite{marel2001tran, marel2004opti, koyama2002jose}. The frequency of the latter is in the mid-infrared range, it is of the same order of the superconducting gap, and it correlates with the critical temperature \cite{hirata2012corr}. Moreover, the JPW has recently become of great interest for light-induced superconductivity \cite{fausti2011} and for nonlinear propagation of mixed plasma--electromagnetic waves in the form of Josephson plasma solitons \cite{dienst2013}.

In this work, we consider a small particle made of a multilayered superconductor, which is modeled as a macroscopic medium with an anisotropic dielectric tensor: as a benchmark example leading to a simple solution, we take the particle to have a spherical shape. In the quasistatic limit, we derive a general analytic formula which provides the eigenmodes corresponding to \emph{Josephson surface plasmons} (JSPs), fully confined at the particle, in the whole frequency range and for any multipolar symmetry. They represent the generalization of the well-known surface plasmons of a metallic nanoparticle, but with a far more complex spectrum, due to the presence of longitudinal and transverse JPWs, quasi-2D plasmons, and mutal couplings among them. All these effects are also relevant for the interpretation of experiments with the sphere resonance method.

The paper is organized as follows. In Sec.~\ref{sec:bulk} we summarize the properties of \emph{bulk} electromagnetic modes of cuprate superconductors. In Sec.~\ref{sec:surface} we treat surface modes that originate at dielectric--superconductor interfaces, first for the planar case and, then, for the fully confined geometry of a small spherical particle. Calculations are carried out for a $n = 2$ multilayered superconductor. We also evaluate optical extinction and electron energy-loss spectra, the latter being a probe of surface plasmons of any multipolar symmetry. Finally, in Sec.~\ref{sec:conclusion} we discuss our results in view of experiments on actual samples with more realistic shapes and we formulate some concluding remarks. Details on the calculations for bulk electromagnetic modes and localized surface plasmons in the spherical geometry are contained in Appendices~\ref{app:TM} and \ref{app:sphere}, respectively.

\section{Bulk modes}\label{sec:bulk}

As a starting point, we discuss the electromagnetic modes of bulk cuprate superconductors. The characteristic frequency of the \emph{acoustic} Josephson plasma wave is generally in the terahertz range\cite{noh1990far,sphere1,sphere2} (however for Bi$_{2}$Sr$_{2}$CaCu$_{2}$O$_{8+x}$ it is in the gigahertz range\cite{wj1_bscco}), whereas the characteristic frequency of the \emph{optical} JPW varies with different multilayered superconductors, ranging from \SI{30}{cm^{-1}} (in SmLa$_{0.8}$Sr$_{0.2}$CuO$_{4-x}$\cite{wj2_slsco}) up to \num{400}--\SI{500}{cm^{-1}} in YBa$_{2}$Cu$_{3}$O$_{{7-x}}$ and Bi$_{2}$Sr$_{2}$CaCu$_{2}$O$_{8+x}$\cite{wj2_ybco1,wj2_ybco2,wj2_bscco1,wj2_bscco2}. In-plane quasi-2D plasmons are characterized by frequencies of the order of the electronvolt\cite{bozovic,romero}. The wavelengths corresponding to all these excitations result significantly larger than the typical size of the unit cell, which is of the order of a few \AA{}ngstrom\cite{plakida}. For this reason, the optical response of \highTc{} cuprate superconductors can be suitably modeled
with macroscopic electrodynamics in the long-wavelength approximation, as commonly done in the literature \cite{helm2002opti,report,rakhmanov2010laye}. Moreover, when we are interested in the low-frequency range of the spectrum, the wavelengths are larger than several 10 to \SI{100}{\upmu{}m} and the electromagnetic response is very weakly sensitive to the presence of local inhomogeneity at the microscale.

The dispersion of transverse magnetic (TM) modes in a single-layer superconductor (such as Bi$_{2}$Sr$_{2}$CuO$_{6+x}$) can be derived from the microscopic Lawrence-Doniach model\cite{LD} as shown in App.~\ref{app:TM}. The dispersion relation, yielding the frequency $\w = \w(k_x,k_z)$ as a function of the wavevector $\vec{k} = (k_x,0,k_z)$, assumes the form
\begin{widetext}
\begin{equation}\label{eq:TM_gen}
\cosh(\alpha s) - \cos(k_z s) =
\frac{s\,\sinh(\alpha s)}{2 \alpha}
\left\{\frac{\epsoo - \epsab(\w)}{\epsoo}\,\alpha^2
+ \frac{4k_x^2 \sinsquared}
{s^2\left[k_x^2 - \frac{w^2}{c^2}\epsc(\w)\right]}\,
\frac{\epsoo - \epsc(\w)}{\epsoo}\right\},
\end{equation}
\end{widetext}
where $s$ is the interplane distance, $\epsoo$ the high-frequency dielectric constant, and $\alpha^2 = k_x^2 - \epsoo\w^2 / c^2$. The $c$-axis is directed as $\zv$. The dielectric function
\begin{equation}\label{eq:epsab}
\epsab(\w)=\epsoo\left(1-\frac{\wpl^2}{\w^2}\right)
\end{equation}
is a Drude function which describes the in-plane response of the charge carriers (for the moment, dissipation is neglected). The screened in-plane plasma frequency $\wpl$ depends on the average density of the carriers and is generally of the order of the electronvolt \cite{bozovic,romero}. The dielectric function $\epsc(\w)$, on the other hand, takes into account interplane tunneling effects due to Josephson coupling. It presents a simple Drude-like behavior with the Josephson plasma frequency $\wj$:
\begin{equation}
\epsc(\w)=\epsoo\left(1-\frac{\wj^2}{\w^2}\right).
\end{equation}
The Josephson plasma frequency is defined as $\wj = \sqrt{8\pi J_0 e s/(\hbar \epsoo)}$ ($J_0$ is the interlayer Josephson critical current) and it is related to the $c$-axis London penetration depth by $\wj = c / (\sqrt{\epsoo}\lambda_c)$.

In the long-wavelength approximation, the dispersion relation in Eq.~\eqref{eq:TM_gen} reduces to
\begin{equation}\label{eq:TM}
\frac{k_x^2}{\epsc(\w)} + \frac{k^2_z}{\epsab(\w)} =
\frac{\w^2}{c^2},
\end{equation}
in agreement with Ref.~\onlinecite{helm2002opti}. TM modes can be interpreted as retarded Josephson plasma waves along the $c$-axis coupled to quasi-2D plasmons on the conductive planes. 

In multilayered superconductors with two planes per unit cell (such as Bi$_{2}$Sr$_{2}$CaCu$_{2}$O$_{8+x}$), however, nonequivalent junctions with inter- and intra-bilayer coupling mechanisms are present, resulting in two JPW frequencies $\wj[1]$ and $\wj[2]$, respectively. Moreover, a transverse-optical Josephson plasma mode arises, whose frequency $\wt$ is intermediate between $\wj[1]$ and $\wj[2]$. The situation has been extensively studied in Refs.~\onlinecite{marel2001tran, marel2004opti, koyama2002jose}, leading to the result that the $c$-axis dielectric function has to be modified in the form
\begin{equation}\label{eq:epsc_bi}
\epsc(\w) = \epsoo\frac{\left(\w^2 - \wj[1]^2\right) \left(\w^2 - \wj[2]^2\right)}
{\w^2\,\left(\w^2 - \wt^2\right)}.
% = \epsoo
%\left[\frac{\xt}{1 - \frac{\wj[1]^2}{\w^2}} + \frac{1 - \xt}{1 - \frac{\wj[2]^2}{\w^2}}\right]^{-1}.
\end{equation}
The lower Josephson frequency $\wj[1]$ can be more than two orders of magnitude smaller than $\wpl$, resulting in very strong anisotropy effects, which are crucial for the electrodynamic response in the superconducting phase.

In addition to TM modes, cuprate superconductors also possess transverse electric (TE) modes, characterized by the dispersion relation
\begin{equation}\label{eq:TE}
k_x^2 + k^2_z = \frac{\w^2}{c^2}\epsab(\w).
\end{equation}
Notably, TE modes are less interesting for the study of interlayer Josephson coupling in \highTc{} superconductors, since they are not affected by the out-of-plane dynamics of the condensate.

\begin{figure}
\includegraphics[scale=0.95]{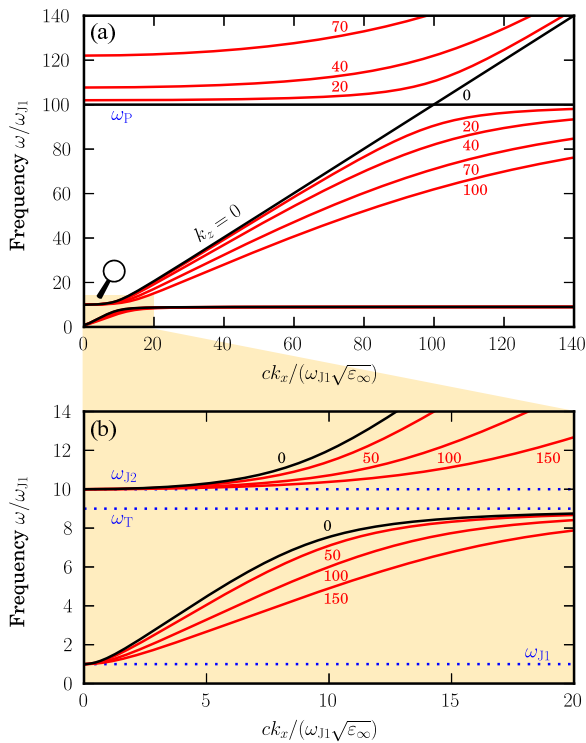}
\caption{\textit{(color online)} (a) Dispersion of the optical TM modes of a bilayered superconductor, as a function of the $ab$-plane wavector $k_x$, normalized to $\sqrt{\epsoo}\wj[1]/c$. Each curve corresponds to a value of the $c$-axis wavector $k_z$ (with the same normalization), indicated by the label. (b) Close-up of the low-frequency and long-wavelength region of the dispersion. Parameters are: $\wt = 9\wj[1]$, $\wj[2] = 10\wj[1]$, and $\wpl = 100\wj[1]$.}\label{fig:bulk}
\end{figure}

TM modes, on the other hand, manifest a clear signature of Josephson tunneling effects. For instance, the dispersion of TM modes resulting from Eq.~\eqref{eq:TM} is represented in Fig.~\ref{fig:bulk} for the case of a $n = 2$ multilayered superconductor, showing distinctive features over a large range of frequencies. On the scale of $\wpl$, the plasmon-polariton dispersion closely resembles that of quasi-2D acoustic plasmons in a layered electron gas with no interplane tunneling \cite{fetter,bill2003elec}. At lower frequencies, however, the acoustic character of the quasi-2D plasmons is lost and a stop-band opens below the lower Josephson plasma frequency $\wj[1]$, as a consequence of interlayer Josephson tunneling. Moreover, a second stop-band is present at the intermediate range of frequencies between $\wt$ and $\wj[2]$, due to the effect of the transverse-optical JPW. The stop-band can be revealed as a \emph{Reststrahl} feature in the reflectance spectrum \cite{munzar,hirata2012corr}.

In the long-wavelength regime under consideration, the electromagnetic response of bulk cuprate superconductors can be treated in a more compact form with the use of the uniaxial dielectric tensor
\begin{equation}\label{eq:diel_tens}
\epst(\w) = \left[\begin{array}{ccc}
\epsab(\w) & 0 & 0 \\
0 & \epsab(\w) & 0 \\
0 & 0 & \epsc(\w)
\end{array}\right],
\end{equation}
with the dielectric functions defined above. Transverse magnetic and transverse electric modes in Eqs.~\eqref{eq:TM} and~\eqref{eq:TE} play the role of extraordinary and ordinary waves, respectively. The validity of Eq.~\eqref{eq:diel_tens} follows from the derivation of Eq.~\eqref{eq:TM} as the long-wavelength approximation of the microscopic dispersion relation \eqref{eq:TM_gen}. According to the macroscopic approach, thus, cuprates can be modeled as anisotropic materials where the strong anisotropy combined with the presence of different Josephson plasma frequencies leads to highly structured optical response over several decades of frequencies.
Notice that this macroscopic description holds irrespectively of correlation effects, and it remains valid even in the presence of local (sub-micron) inhomogeneities, as discussed above.

\section{Surface modes}\label{sec:surface}

In this Section, we study the spectrum of surface plasmon modes that arise when cuprate superconductors are spatially confined. We suppose that the superconducting region is immersed in a material with dielectric constant $\epsd$ and adopt the quasistatic approximation, which is valid when the size of the system is smaller than the wavelength of light under consideration. The requirements on the system size for the quasistatic approximation to hold are different for different spectral regions. For the acustic and optical JPWs (with characteristic frequencies in the range \num{1}--\SI{15}{THz}, as observed in Sec.~\ref{sec:bulk}) particles up to a few $\upmu$m diameter are generally sufficient. For in-plane plasmons ($\wpl\sim \SI{1}{eV}$), the system size is restricted to less than a few \SI{100}{nm}.

In the macroscopic approach, the problem of calculating surface plasmons in confined cuprate superconductors reduces to the more general problem of solving the electrostatic problem in confined geometry for a strongly anisotropic material with a dielectric tensor such as that in Eq.~\eqref{eq:diel_tens}. A similar problem has been addressed, limited to the dipolar approximation, in the context of optical scattering from anisotropic dielectric particles\cite{bohren}. Here, we present a complete analysis of surface plasmons for the introductory example of a planar surface and for the benchmark situation of a spherical particle. 

\subsection{Plasmons in the half-space}

As a first preliminary example, we calculate the characteristic frequencies of surface modes on a dielectric--superconductor interface perpendicular to the superconductor $c$-axis (e.g., along the $z = 0$ plane). The basic procedure is to solve the anisotropic equation for the scalar potential $\nabla\cdot(\epst\,\nabla\phi) = 0$ inside and outside the boundary with the proper continuity conditions at the interface. The potential in the two regions $z > 0$ (dielectric) and $z < 0$ (superconductor) can be written in the form
\begin{equation}
\phi(\rv) = e^{i k x}\begin{cases}
A\,\exp(-kz), \qquad & z > 0;\\
B\,\exp\left(\sqrt{\frac{\epsab}{\epsc}}kz\right), & z < 0.
\end{cases}
\end{equation}
The application of boundary conditions leads to the characteristic equation
\begin{equation}\label{eq:planar}
\sqrt{\epsab(\w)\,\epsc(\w)} = \epsd,
\end{equation}
whose solutions represent the frequencies of instantaneous (quasistatic) surface plasmons over the planar interface. Notice that, in the isotropic limit, the well-known condition $\varepsilon(\w) = - \epsd$ is recovered.

The solutions are constrained to lie in the spectral regions where the $\epsc(\w)$ and $\epsab(\w)$ have the same sign. In particular, as a consequence of strong anisotropy, for cuprate superconductors the frequencies of planar surface plasmons are closer to the plasma frequencies than in the isotropic case. This phenomenon is evident for single-layer superconductors in the $\epsoo = \epsd = 1$ situation, by comparing the anisotropic solution of Eq.~\eqref{eq:planar},
\begin{equation*}
\w_{\text{PL}} = \frac{\wj[]}{\sqrt{1 + \wj^2/\wpl^2}} \lesssim \wj[],
\end{equation*}
with its well-known isotropic counterpart $\w_{\text{PL}} = \wj[] / \sqrt{2}$.

For a $n = 2$ cuprate superconductor, application of Eq.~\eqref{eq:planar} leads to two different solutions for istantaneous surface plasmons, which are located in close proximity to the Josephson plasma frequencies $\wj[1]$ and $\wj[2]$.

\subsection{Plasmons in spherical geometry}

As we have already stressed, in this work we are mostly concerned with the effect of spatial confinement for plasmonic modes in high-\Tc{} superconductors. As a benchmark system fully confined in three dimensions, we consider a spherical particle made of a multilayered superconductor. The electrostatic problem with the proper boundary conditions on the spherical surface is solved by means of a coordinate transformation to an ellipsoidal geometry, where the usual isotropic Laplace equation is recovered, as elaborated in App.~\ref{app:sphere}. As a result, we are led to the characteristic equation
\begin{equation}\label{eq:characteristic}
\epsc(\w)\,\kappa\,\frac{d}{d\kappa}\Plm(\kappa)
 + \epsd\,(l + 1) \Plm(\kappa) = 0,
\end{equation}
with $\kappa = [1 - \epsc(\w) / \epsab(\w)]^{-\frac{1}{2}}$ and $\Plm$ the associated Legendre polynomials. This equation is a central result of the present work. In particular, we stress that it holds for any uniaxial material: the specific properties of the solutions are entirely dependent  on the behavior of the dielectric functions $\epsab(\w)$ and $\epsc(\w)$. For the case of multilayered superconductors, its solutions represent a spectrum of surface excitations that are localized close to the superconducting particle and which we call \emph{Josephson surface plasmons} (JSPs). Each surface mode is identified by the ``azimuthal'' quantum number $l$ and the ``magnetic'' quantum number $m$. Each $m>0$ mode is degenerate with its $-m$ counterpart; however, at variance from the case of spherical particles made of isotropic materials, modes with the same $l$ and different $m$ are nondegenerate. Equation~\eqref{eq:characteristic} reduces to the well-known result for spheres of isotropic media in the limit situation $\epsab = \epsc$, as shown in App.~\ref{app:sphere}.

\begin{figure}[tb]
\includegraphics{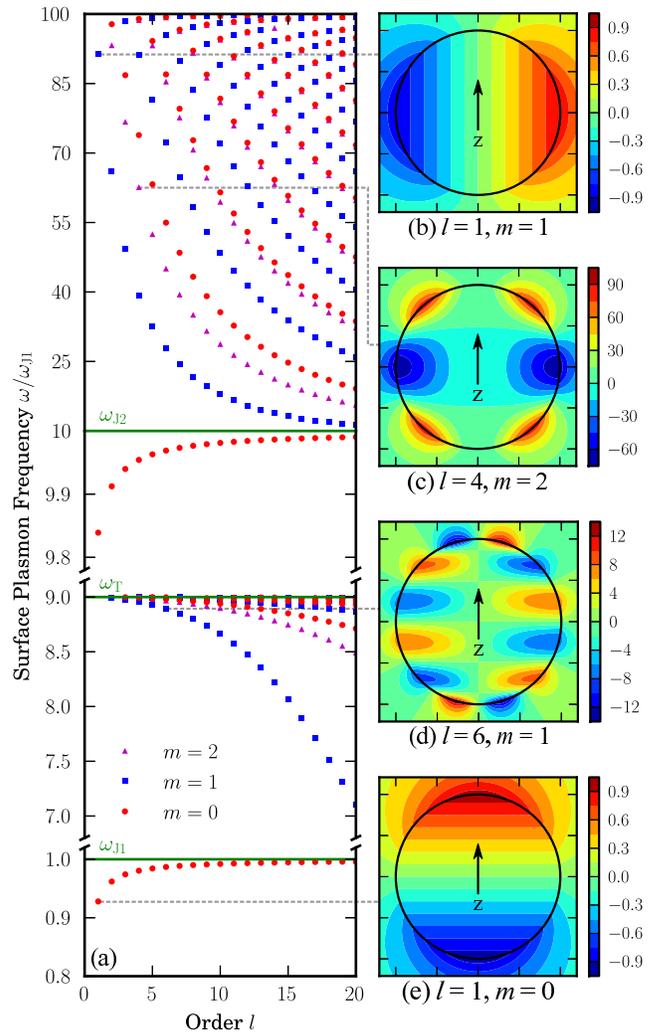}
\caption{\textit{(color online)} Quasistatic Josephson surface plasmons in a spherical particle made of a $n = 2$ multilayered superconductor. (a) The modal frequencies, as a function of the azimuthal quantum number $l$ and for three values of the magnetic quantum number $m = 0,1,2$. Notice the different scales in the frequency ranges of the $\w$-axis. Parameters are: $\wt = 9\wj[1]$, $\wj[2] = 10\wj[1]$, $\wpl = 100\wj[1], \epsoo = 10$ and $\epsd = 1$. (b--e) Potential maps of the modes indicated by the dashed lines. Potential is shown on a cut along the $xz$ plane including the center of the sphere, with the $c$-axis oriented as $\zv$.}\label{fig:modes}
\end{figure}

The spectrum for the case of $n = 2$ multilayered superconductor is shown in Fig.~\ref{fig:modes}(a) as a function of the azimuthal number $l$ for some values of the magnetic number $m$ and it spans a large range of frequencies up to the in-plane plasma frequency $\wpl$. In particular, several $m = 0$ modes lie entirely inside the stop-bands, as their charge oscillations are mostly related to Josephson plasma waves along the $c$-axis. In the remaining regions of the spectrum, a large number of closely spaced Josephson surface plasmons gives rise to a quasi-continuum of excitations \footnote{Figure \ref{fig:modes} for simplicity shows only the modes up to $m = 2$; however a large number of additional modes with $m > 2$ are present with a similar dispersion pattern.}.

When $l$ tends to infinity for fixed $m$, Eq.~\eqref{eq:characteristic} reduces to the corresponding relation for a planar surface, i.e., Eq.~\eqref{eq:planar}. Indeed, with the incresing of $l$, sphere modes view the surface as a planar one\footnote{The convergence to the planar-surface-plasmon frequency for $l \to \infty$ is observable even for an \emph{isotropic} metallic sphere.}. For this reason, the true watersheds between different regions of the excitation spectrum are the frequencies of the surface plasmons on a planar interface, defined as the solutions of Eq.~\eqref{eq:planar}. As we noticed before, these frequencies are very close to the Josephson plasma frequencies $\wj[1,2]$ and, in Fig.~\ref{fig:modes}, they are practically indistinguishable from them. 

An important family of Josephson surface plasmons is that with $l = 1$ (dipolar modes). Here, Eq. \eqref{eq:characteristic} reduces to the isotropic forms
\begin{align*}
\epsc(\w) + 2\epsd &= 0 \qquad & (m=0),\\
\epsab(\w) + 2\epsd &= 0 & (m=1),
\end{align*}
implying that the $m=0,1$ JSPs have the same frequencies of those of isotropic spheres with dielectric functions equal to $\epsc(\w)$ and $\epsab(\w)$, respectively. The $m = 1$ mode at $\omega/\wj[1]=91.3$ can be interpreted as a quasi-2D plasmon confined in the $ab$-planes, as confirmed by the potential profile in Fig.~\ref{fig:modes}(b), which reveals an oscillating electric field perpendicular to the $c$-axis inside the particle. Analogously, the two $m = 0$ modes at $\omega/\wj[1]=0.93$ and $9.84$ can be considered longitudinal JPWs confined along the $c$-axis, in agreement with Fig.~\ref{fig:modes}(e). This is consistent with the fact that the $m=0$ modes lie within the stop-bands below the Josephson frequencies $\wj[1]$ and $\wj[2]$.

For the generic modes with $l > 1$ this interpretation is not valid anymore, since $\epsc(\w)$ and $\epsab(\w)$ are coupled together in Eq.~\eqref{eq:characteristic}; for instance, in the case $l = 2, m = 0$, the characteristic equation becomes
\begin{equation}
\frac{1}{\epsd} + \frac{1}{\epsc(\w)} + \frac{1}{2\epsab(\w)} = 0.
\end{equation}
The $l > 1$ plasmonic modes do not have a specific Josephson or quasi-2D character, but they present confinement both along the optical axis and perpendicular to it, as shown by the potential plots in Fig. \ref{fig:modes}(c--d). This can be interpreted as a phenomenon of \emph{mixing} of JPWs and quasi-2D plasmons, which is a peculiar feature of the multipolar surface modes of confined layered superconductors.

\subsection{Probing of surface plasmons}

Measuring Josephson surface plasmons requires employing an external probe and investigating, for instance, the extinction cross section or the electron energy loss (EEL) spectrum. The theoretical calculation of these quantities depends on the \emph{multipolar polarizabilities} $\alpha_{lm}(\w)$. Several results for spheres of isotropic media can be directly generalized to the anisotropic case, as long as the polarizabilities $\alpha_{lm}$ are redefined as
\begin{equation}\label{eq:alpha}
\alpha_{lm}(\w) = \frac{\epsc(\w)\,\kappa\,\frac{d}{d\kappa}\Plm(\kappa) -
\epsd l\,\Plm(\kappa)}
{\epsc(\w)\,\kappa\,\frac{d}{d\kappa}\Plm(\kappa) + \epsd(l+1)\,\Plm(\kappa)}R^3
\end{equation}
[$\kappa$ is defined as in Eq. \eqref{eq:characteristic} and $R$ is the sphere radius.] In the limit of zero dissipation, $\alpha_{lm}$ behaves like a delta function centered on the zeros of Eq. \eqref{eq:characteristic}, i.e, the JSP frequencies. However, due to the presence of noncondensed carriers, a certain amount of dissipation occurs and it is responsible for linewidth broadening of the peaks. Moreover, a radiative broadening of JSPs is expected when going beyond the quasistatic approximation.

\begin{figure}
\includegraphics{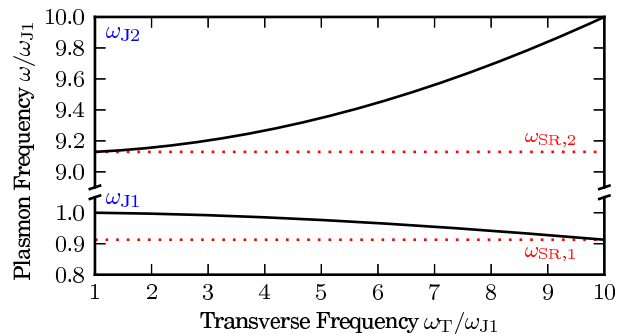}
\caption{\textit{(color online)} The frequencies of the two $l = 1$, $m = 0$ JSPs of the same particle as in Fig. \ref{fig:modes}, versus the transverse-optical Josephson frequency $\wt$. The (red) dotted lines indicate the frequencies calculated with Eq.~\eqref{eq:sr}.}
\label{fig:dipole}
\end{figure}

Once redefined the multipolar polarizabilities as in Eq.~\eqref{eq:alpha}, the extinction cross section for an incident plane wave in the dipolar approximation is given by \cite{bohren}
\begin{equation}
C_{\text{ext}} = \frac{4\pi\w}{c}\Im\left(\alpha_{10}\cos^2\thetain
+ \alpha_{11} \sin^2\thetain\right),
\end{equation}
where $\thetain=\arccos(\ver{E}\cdot\ver{c})$ is the polar angle of the electric field. Indeed, scattering from small objects is dominated by dipolar modes, while higher-order modes will become manifest for larger particles \footnote{The same dipolar or multipolar coupling gives rise to radiative broadening when treated in the fully retarded approach. The radiative linewidth is $\propto R^{2l+1}$ (Ref. \onlinecite{colas}), so that the effect is negligible for small particles for which the quasistatic approximation applies.}.

These results are relevant for sphere-resonance experiments. Usually, the absorption resonances are associated to the frequencies
\begin{equation}\label{eq:sr}
\w_{\text{SR},\alpha}=\w_{\mathrm{J}\alpha}\sqrt{\frac{\epsoo}{\epsoo+2\epsd}}, \quad \alpha=1,2.
\end{equation}
However, the frequencies of the dipolar $l = 1$, $m=0$ JSPs obtained from Eq. \eqref{eq:characteristic} generally differ from the expressions \eqref{eq:sr} as an effect of the transverse-optical JPW. This can be seen in Fig. \ref{fig:dipole}, where the JSP frequencies are plotted as a function of $\wt$. In the limit $\wt\to\wj[2]$ the lowest JSP mode tends to $\w_{\mathrm{SR},1}$, while the difference is maximum for the upper JSP mode. In the limit $\wt\to\wj[1]$, on the other hand, the situation is reversed. These results show that a careful analysis of sphere-resonance experiments should be undertaken in order that $\wj[1]$, $\wj[2]$ and $\wt$ are simultaneously determined.

\begin{figure}[t]
\includegraphics{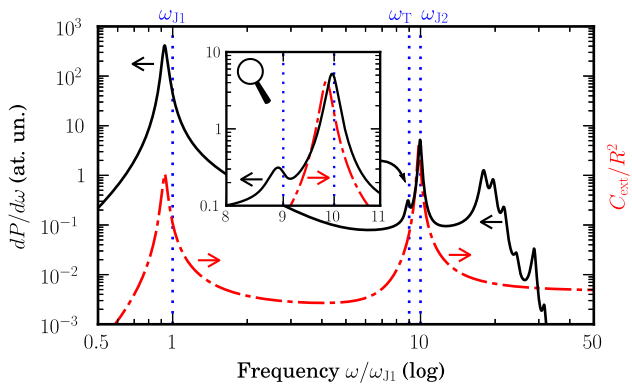}
\caption{\textit{(color online)} Dark solid line: the EEL spectrum of a \SI{300}{eV} electron moving along the $c$-axis at grazing incidence ($b = 1.02R$) on a $R = \SI{400}{nm}$ sphere of a bilayered superconductor. Dashed line: the extinction cross section, for a field polarized along the $c$-axis. Here, $\wj[1] = \SI{10}{meV}$ and the parameters are as in Fig. \ref{fig:modes} (vertical dotted lines). In-plane dissipation is modeled assuming $\Im\epsab = 10\wj[1]\wpl^2/\w^3$, whereas inter- and intra-bilayer tunneling are supplemented with the additional conductance $4\pi\sigma_{\text{c1}} = \wj[1]/2$ and $4\pi\sigma_{\text{c2}} = \wj[2]/4$, respectively. Inset: close-up around $\wt$ and $\wj[2]$.}
\label{fig:eels}
\end{figure}

At variance from extinction spectroscopy, electron energy loss spectroscopy allows to excite the full spectrum of multipolar JSPs, as shown by the EEL probability function for an electron with impact parameter $b > R$ (moving outside the particle along the $c$-axis)\cite{eels1,*eels2}
\begin{equation*}
\frac{dP}{d\w} =
\frac{4}{\pi v^2 R^2} \sum_{l,m} M_{lm}
\left(\frac{\w R}{v}\right)^{2l}K^2_m\left(\frac{\w b}{v}\right)
\Im\left(\alpha_{lm}\right),
\end{equation*}
where $K_m(x)$ are the modified Bessel functions, $v$ is the electron velocity, and $M_{lm} = (2-\delta_{m0})/[(l+m)!(l-m)!]$. The EEL spectrum for a \SI{300}{eV} electron is represented by the solid curve in Fig. \ref{fig:eels} and compared with the extinction cross section, indicated by the dashed curve. The effect of higher-order Josephson surface plasmons is particularly evident in the region around $\wt$ and $\wj[2]$. The peak in proximity of $\wj[2]$ is blue-shifted with respect to the extinction cross section, as a consequence of the increasing weight of the $l > 1$ higher-frequency modes (see Fig. \ref{fig:modes}); moreover, an additional peak arises just below $\wt$, due to the excitation of the corresponding modes of nondipolar symmetry.

\section{Discussion and conclusions}\label{sec:conclusion}

The present treatment assumes a spherical geometry for the superconducting particle, which is appropriate as a benchmark example but it is hard to realize experimentally with ceramic compounds. Other shapes are more suitable to fabrication with top-down nanotechnology approaches. For example, epitaxial films of cuprate superconductors could be patterned in the shape of small discs, lying on a dielectric substrate and with the height oriented along the superconductor $c$-axis. For a typical thickness of cuprate films around \num{100}--\SI{300}{nm}\cite{crisan,ooi,palau}, discs with an aspect ratio of a few units (and, so, a diameter up to \SI{1}{\upmu{}m}) are fully compatible with the quasistatic approximation in the terahertz range ($\lambda \sim \SI{300}{\upmu{}m}$). The patterning can be realized by using optical\cite{crisan} or electron-beam lithography\cite{bozovic_pattern} followed by ion milling or else using focused ion beam lithography\cite{ooi,palau}. In order to increase the signal to noise ratio, several discs can be arranged together in a periodic configuration. Since the electrostatic field is strongly localized around the particle [the potential decays as $r^{-(l+1)}$, see Eq.~\eqref{eq:field_out} in App. \ref{app:sphere}], a separation distance of a few times the diameter should be enough to rule out overlapping effects for the fields. This is analogous to the spectroscopy of surface plasmons in usual metallic nanoparticles, which are commonly prepared in dense --- ordered or disordered --- arrangements\cite{maier,gaponenko,review_plasmons}.

Despite the exact frequencies of Josephson plasmons being modified, we believe that the general qualitative features of surface plasmons in cylindrical geometry remain similar to those presented in this work. More detailed calculations can be performed with a full computational approach, either by adapting existing numerical methods or by expanding the eigenmodes of an arbitrary particle in the basis consisting of the Josephson surface plasmons of the spherical geometry.

To sum up, we have analyzed the effect of spatial confinement onto the optical response of high-\Tc{} cuprate superconductors, giving rise to a discrete spectrum of collective excitations of the condensate, which we call \emph{Josephson surface plasmons}. The  spectrum of surface plasmons contains a rich variety of dipolar and multipolar modes, which span a wide frequency range up to the in-plane plasma frequency. For multilayered superconductors, spatial confinement leads to delicate coupling effects between acoustic and optical Josephson plasma waves. Confined surface plasmons can be probed by optical absorption or by EEL spectroscopy, the latter being sensitive to modes of any symmetry.

The present approach, in addition, is suitable to study localized surface excitations for arbitrary uniaxial materials. It can also be extended to superconducting particles of non-spherical shapes, in analogy with surface plasmons in metal nanoparticles, but with the additional effects of strong dielectric tensor anisotropy. We believe that the study of Josephson plasmons via their size confinement is a promising new route to explore phenomena associated with interlayer tunneling in cuprates and, possibly, their relation to high-\Tc{} superconductivity.

\section*{Acknowledgment}
One of us (LCA) is grateful to Lucia Bossoni for introducing him to the problem, and for many precious discussions.

\appendix

\section{Electromagnetic modes of a single-layer superconductor}\label{app:TM}

In this Appendix, we sketch the derivation of the electromagnetic dispersion for transverse magnetic (TM) modes in a single-layer cuprate superconductor, reported in Eq.~\eqref{eq:TM_gen}. We model the superconductor as a stack of two-dimensional conductive planes separated by the distance $s$, located at the positions $z = ns$ (with integer $n$), and immersed in a background material with dielectric constant $\epsoo$.

As a consequence of interlayer Josephson tunneling, the average current density flowing from the $n^{\text{th}}$ to the $(n+1)^{\text{th}}$ plane can be written
as\cite{helm2002opti,report}
\begin{equation}\label{eq:rcsj}
J_z\llp = - J_0 \sin[\xi\llp] +
i\frac{\hbar\w\sigmaperp}{2 e s}\,\xi\llp,
\end{equation}
where $\xi\llp$ is the gauge invariant phase difference between the superconducting layers. The out-of-plane conductivity $\sigmaperp$ accounts for the tunneling of noncondensed electrons and it is responsible for dissipation. Since we are interested in the dispersive properties of the modes, for the moment we consider the limit situation $\sigmaperp \to 0$. Neglecting the effect of the breaking of charge neutrality\cite{report} (which is appropriate in the long wavelength approximation), the phase difference is proportional to the averaged $z$-component of the electric field along the junction
\begin{equation}
\xi\llp = -i\frac{2e s}{\hbar\w}E_z\llp.
\end{equation}
The layered structure of the superconductor can be viewed as a superlattice. Thus, as a consequence of the Bloch--Floquet theorem, the in-plane electric field on the $n^{\text{th}}$ plane can be written as $E_x(z = ns) = \mathcal{E}_n e^{ink_z s}$.

TM modes are characterized by the non-null components $E_x,E_z$, and $H_y$ of the electric and magnetic fields, respectively. Their dispersion relation can be calculated from the vector equation
\begin{equation}
\nabla(\nabla\cdot\Ev) - \nabla^2\Ev =
\epsoo\frac{\w^2}{c^2}\Ev + i\,\frac{4\pi\w}{c^2}\vec{J}.
\end{equation}
Upon performing the Fourier transform with respect to $x$, the $x$-component of the vector equation becomes
\begin{equation}\label{eq:Ex}
\left[\frac{\partial^2}{\partial z^2} - \alpha^2\right]E_x =
\sum_n \left[ik_x \frac{4\pi}{\epsoo}\,\rho_n\td - i\,\frac{4\pi\w}{c^2}J_x\td\right]\delta(z - ns),
\end{equation}
with $\alpha^2 = k_x^2 - \epsoo\w^2 / c^2$ and $\rho_n\td$ the local charge density on the $n^{\text{th}}$ plane. The in-plane current on the $n^{\text{th}}$ plane follows the ordinary Drude behavior (we neglect dissipation effects)
\begin{equation}
J_x\td = i\frac{s \epsoo}{4\pi\w}\wpl^2\, E_x,
\end{equation}
whereas the charge density can be expressed as a function of the in-plane electric field $E_x$ by applying the continuity relation
\begin{equation*}
i\w\rho_n\td = ik_xJ_x\td + J_z\llp - J_z\llm,
\end{equation*}
the linearized approximation of the Josephson current in Eq.~\eqref{eq:rcsj}, and Maxwell equations. The result is
\begin{equation*}
\rho_n\td = k_x\left\{\frac{J_x\td}{\w} +
i\,\frac{\sinsquared}{\pi s \left(k_x^2 - \epsc w^2 / c^2\right)}\left[\epsoo - \epsc\right] E_x\right\}.
\end{equation*}

Equation \eqref{eq:Ex} has the form
\begin{equation*}
\left[\partial^2 / \partial z^2 - \alpha^2\right]E_x =
- \delta(z-z_0),
\end{equation*}
whose solution is\cite{fetter} $E_x(z) = (2\alpha)^{-1}e^{-\alpha|z-z_0|}$. At this point, we can follow the procedure in Ref.~\onlinecite{fetter} and obtain the dispersion relation in Eq.~\eqref{eq:TM_gen} of Sec.~\ref{sec:bulk} from the result
\begin{equation}
\sum_n e^{-(ik_z n + \alpha|n|)s} = \frac{\sinh(\alpha s)}{\cosh(\alpha s) - \cos(k_z s)}.
\end{equation}
The dispersion relation for transverse electric (TE) modes can be obtained with a similar procedure. TE modes have a null $z$-component of the electric field, so they are not affected by the Josephson tunneling current in Eq.~\eqref{eq:rcsj}.

%%%%%%%%%%%%%%%%%%%%%%%%%%%%%%%%%%%%%%%%%%%%%%%%%%%%%%%%%%%%%%%%%%%%%%%%%%%%%%%%%%
\section{Surface plasmons of a spherical particle in the presence of anisotropy}\label{app:sphere}

In this Appendix, we consider a spherical particle with radius $R$, filled by an anisotropic material with a diagonal dielectric tensor such as that in Eq.~\eqref{eq:diel_tens}. We solve the electrostatic problem inside and outside the particle and, from the boundary conditions at the interface, we derive the characteristic equation for surface plasmon modes [Eq.~\eqref{eq:characteristic}].

In spherical coordinates, the electrostatic potential in the region outside the particle has the form\cite{khurgin}
\begin{equation}\label{eq:field_out}
\phi_{\text{out}}(\rv) = \sum_{lm} A_{lm} r^{-(l+1)}\Plm(\cos\theta)\left\{\begin{array}{c}
\cos(m\varphi) \\
\sin(m\varphi) \end{array}\right.,
\end{equation}
where $\Plm$ are the associated Legendre polynomials and the $\varphi$-dependent part of the potential varies for even and odd modes, respectively. The effect of anisotropy inside the sphere can be taken into account by means of the coordinate transformation $\LL$, defined as
\begin{equation*}
\left[\begin{array}{c}x \\ y \\ z \end{array}\right] \to
\left[\begin{array}{c}\xt \\ \yt \\ \zt \end{array}\right] =
\LL \left[\begin{array}{c}x \\ y \\ z \end{array}\right] =
\left[\begin{array}{c}x \\ y \\ \sqrt{\frac{\epsab}{\epsc}}z \end{array}\right].
\end{equation*}
We introduce the potential $\phit_{\text{in}}$ in the transformed space, which is connected to the potential inside the particle by the relation $\phit_{\text{in}}(\LL\rv) = \phi_{\text{in}}(\rv)$. The transformation $\LL$ is chosen so that the \emph{anisotropic} equation $\nabla\cdot(\epst\,\nabla\phi_{\text{in}}) = 0$ reduces to the \emph{isotropic} Laplace equation in the transformed space
\begin{equation}
\nablat^2\phit_{\text{in}}(\rt) = 0;\qquad \rt = \LL\rv.
\end{equation}
Notice that the coefficient $\sqrt{\epsab / \epsc}$ of the transformation could assume purely imaginary values, implying that the components of $\rt$ could be purely imaginary, as well. This is due to the fact that transformed coordinates are just a mathematical device without an explicit physical meaning; measurable quantities, such as the electric field, remain real even in the latter situation.

The linear operator $\LL$ transforms the spherical surface of the particle to a spheroidal surface. The solution of the Laplace equation in spheroidal coordinates can be expanded in the form\cite{spencer}
\begin{equation}\label{eq:field_in}
\phit_{\text{in}}(\rt) = \sum_{lm} B_{lm} \Plm(\cosh\etat)
\Plm(\cos\thetat)\left\{\begin{array}{c}
\cos(m\varphi) \\
\sin(m\varphi) \end{array}\right.,
\end{equation}
where $\etat$ and $\thetat$ are the prolate spheroidal coordinates in the transformed space, defined as in Ref.~\onlinecite{spencer}:
\begin{align*}
\xt &= a\sinh\etat\sin\thetat\cos\varphi, \\
\yt &= a\sinh\etat\sin\thetat\sin\varphi, \\
\zt &= a\cosh\etat\cos\thetat,
\end{align*}
with $a = R\sqrt{\epsab / \epsc - 1}$. The spheroidal surface in transformed space that corresponds to the spherical surface in real space is given by the condition $\etat = \etat_0$, with $\tanh\etat_0 = \sqrt{\epsc / \epsab}$. The choice of the factor $a$ has been made so that the coordinates $\theta$ and $\thetat$ coincide on the spheroidal surface.

On the particle surface, the radial component of the external displacement field takes the form
\begin{equation}\label{eq:D_out}
D_{\text{out}}^{\perp} = \sum_{lm} \epsd A_{lm} (l+1) R^{-(l+2)}\Plm(\cos\theta)\left\{\begin{array}{c}
\cos(m\varphi) \\
\sin(m\varphi) \end{array}\right..
\end{equation}
On the other hand, the displacement field inside the particle can be easily related to the gradient of the electrostatic potential in the transformed space
\begin{equation*}
\vec{D}_{\text{in}}(\rv) = - \epst\nabla\phi_{\text{in}}(\rv) = -\epsab\LL^{-1}\nablat \phit_{\text{in}}(\rt).
\end{equation*}
This allows to express the radial component of the internal displacement field on the particle surface as
\begin{multline}\label{eq:D_in}
D_{\text{in}}^{\perp} = -\sum_{lm} \epsc \frac{B_{lm}}{R}\kappa \frac{d}{d\kappa}\Plm(\kappa) \Plm(\cos\theta)\left\{\begin{array}{c}
\cos(m\varphi) \\
\sin(m\varphi) \end{array}\right.,
\end{multline}
with $\kappa = \cosh\etat_0 = [1 - \epsc / \epsab]^{-\frac{1}{2}}$. Boundary conditions are enforced by equating each term in Eqs.~\eqref{eq:field_out} and \eqref{eq:field_in} on the surface ($r = R$ and $\etat = \etat_0$, respectively), and each term of Eqs.~\eqref{eq:D_out} and \eqref{eq:D_in}. The result is the characteristic equation reported in Eq.~\eqref{eq:characteristic} of the main text.

The same Eq.~\eqref{eq:characteristic} reduces to the well-known result for spheres of isotropic materials in the limit case $\epsab = \epsc$, i.e., $\kappa \to \infty$. From the large-argument behavior of associated Legendre polynomials\cite{nist} [$\Plm(\kappa) \sim \kappa^l$], it is easy to see that
\begin{equation*}
\frac{d\Plm(\kappa)}{d\kappa} / \Plm(\kappa) \simeq \frac{l}{\kappa}
\qquad
(\kappa \to \infty),
\end{equation*}
whence we recover the characteristic equation for instantaneous surface plasmons in spheres of isotropic materials\cite{maier,khurgin}
\begin{equation}
l\epsc(\w) + (l+1)\epsd = 0.
\end{equation}

\bibliography{biblio,new}

\end{document}